\documentclass[american,aps,pra,reprint]{revtex4-1}
\usepackage[T1]{fontenc}
\usepackage[latin9]{inputenc}
\usepackage{babel}
\usepackage{amsthm}
\usepackage{amsmath}
\usepackage{amssymb}
\usepackage{graphics,graphicx, epstopdf}
\usepackage{epsfig}
\usepackage[up]{subfigure}
\usepackage[unicode=true,pdfusetitle, bookmarks=true,bookmarksnumbered=false,bookmarksopen=false,
 breaklinks=false,pdfborder={0 0 0},backref=false,colorlinks=false]
 {hyperref}
\hypersetup{colorlinks,linkcolor=myurlcolor,citecolor=myurlcolor,urlcolor=myurlcolor}

\makeatletter
\@ifundefined{textcolor}{}
{%
 \definecolor{BLACK}{gray}{0}
 \definecolor{WHITE}{gray}{1}
 \definecolor{RED}{rgb}{1,0,0}
 \definecolor{GREEN}{rgb}{0,1,0}
 \definecolor{BLUE}{rgb}{0,0,1}
 \definecolor{CYAN}{cmyk}{1,0,0,0}
 \definecolor{MAGENTA}{cmyk}{0,1,0,0}
 \definecolor{YELLOW}{cmyk}{0,0,1,0}
 }
\theoremstyle{plain}

\ifx\proof\undefined
\newenvironment{proof}[1][\protect\proofname]{\par
\normalfont\topsep6\p@\@plus6\p@\relax
\trivlist
\itemindent\parindent
\item[\hskip\labelsep
\scshape
#1]\ignorespaces
}{%
\endtrivlist\@endpefalse
}
\providecommand{\proofname}{Proof}
\fi

\usepackage{times}
\usepackage{txfonts}
\usepackage{braket}
\usepackage{colortbl}
\definecolor{myurlcolor}{rgb}{0,0,0.7}

\makeatother

\providecommand{\theoremname}{Theorem}
\newcommand{\ketbra}[2]{|{#1} \rangle \langle {#2} |}
\usepackage{times}

\usepackage[up]{subfigure}

\begin{document}
\title{Uncertainty Relations for Quantum Coherence}

\author{Uttam Singh}
\email{uttamsingh@hri.res.in}
\affiliation{Harish-Chandra Research Institute, Chhatnag Road, Jhunsi, 
Allahabad 211 019, India}

\author{Arun Kumar Pati}
\email{akpati@hri.res.in}
\affiliation{Harish-Chandra Research Institute, Chhatnag Road, Jhunsi, 
Allahabad 211 019, India}

\author{Manabendra Nath Bera}
\email{manabendra.bera@icfo.es}
\affiliation{ICFO-Institut de Ciències Fotòniques, The Barcelona Institute of Science and Technology, ES-08860 Castelldefels, Spain}


\begin{abstract}
Coherence of a quantum state intrinsically depends on the choice of the reference basis. A natural question to ask is the following: if we use two or more incompatible reference bases, can~there be some trade-off relation between the coherence measures in different reference bases? We show that the quantum coherence of a state as quantified by the relative entropy of coherence in two or more noncommuting reference bases respects uncertainty like relations for a given state of single and bipartite quantum systems. In the case of bipartite systems, we find that the presence of entanglement may tighten the above relation. Further, we find an upper bound on the sum of the relative entropies of coherence of bipartite quantum states in two noncommuting reference bases. Moreover, we provide an upper bound on the absolute value of the difference of the relative entropies of coherence calculated with respect to two incompatible bases.
\end{abstract}

\maketitle
\section{Introduction}
The linearity of quantum mechanics gives rise to the concept of superposition of quantum states of a quantum system and is one of the characteristic properties that makes a clear distinction between the ways a classical and a quantum system can behave. Recently, there has been a considerable effort devoted towards quantifying quantum superposition (quantum coherence) from a resource theoretic perspective \cite{Aberg2006, Baumgratz2014, Girolami14, Bromley2015, Alex15, Xi2015, Winter2015, Fan15, Pinto2015, Du2015, Yao2015, Killoran2015, ZhangA2015, Uttam2015, UttamA2015, Cheng2015, Mondal2015, MondalA2015, Kumar2015, Mani2015, Bu2015, Chitambar2015, StreltsovA2015, ChitambarA2015, StreltsovE2015, Bera2015, Bagan2015, Streltsov2016, Hillery2016, Rana2016, Rastegin2016}.  Also, quantum coherence has been deemed to play a key role in the emerging fields such as quantum thermodynamics \cite{Aspuru13, Horodecki2013, Skrzypczyk2014, Narasimhachar2015, Brandao2015, Rudolph214, Rudolph114, Piotr2015, Gardas2015, Avijit2015, Goold2015} and quantum biology \cite{Abbott2008, Plenio2008,Levi14,Aspuru2009, Lloyd2011, Li2012, Huelga13}. This makes the advancement of a quantitative framework for coherence even more desirable. In particular, a~resource theory of coherence, developed in Ref. \cite{Baumgratz2014}, provides a set of conditions on a real valued function of quantum states for it to be a bona fide quantifier of quantum coherence. This resource theory is based on the set of incoherent operations as the free operations and the set of incoherent states as the set of free states. The set of incoherent states and the set of incoherent operations depend crucially on the choice of basis---the reference basis---and it is determined by the experimental situation at hand.
%
%
Therefore, the quantification of quantum coherence intrinsically depends on the basis that we choose to define the incoherent states and incoherent operations. This seems very unsettling as any physical quantity should not depend on some arbitrary choice of basis. However, the choice of basis is provided naturally by the experimental situation at hand. 

Uncertainty relations play an important role in foundations of quantum mechanics as well as in quantum information science. After the discovery of the Heisenberg uncertainty principle \cite{Heisenberg1927, Kennard1927}, Robertson and Schr\"{o}dinger proved the uncertainty relations for two incompatible observables \cite{Robertson1929, Schrodinger1930}. Recently, two stronger uncertainty relations are proved which go beyond the Robertson-Schr\"{o}dinger uncertainty relation~\cite{Maccone2014}. Undoubtedly, uncertainty relations continue to play pivotal role in quantum theory as well as quantum information science \cite{Busch2007}. In addition to the variance based uncertainty relation, there are entropic uncertainty relations \cite{Birula1975, Deutsch1983, Maassen1988, Berta2010, Coles2015} which capture the notion of uncertainty for two incompatible observables. Entropic uncertainty relations prove to be of fundamental and practical importance in quantum information.

In this work, we explore how the coherence of a given quantum state in one reference basis is restricted by the coherence of the same quantum state in other reference bases for single and bipartite quantum systems. 
For single quantum systems, the trade-offs between the coherences with respect to the mutually unbiased bases (MUBs) have been obtained in Ref. \cite{Cheng2015}. Moreover, the exact complementarity relation for the $l_1$ norm of coherence \cite{Baumgratz2014} of qubit systems with respect to three MUBs is also obtained in Ref. \cite{Cheng2015}.
Here, we go beyond the restriction of the reference bases being MUBs to any set of incompatible bases. Also, we focus on the trade-off relations for bipartite quantum systems and the role of entanglement in these relations.
In particular, we provide lower and upper bounds on the sum of the relative entropies of coherence of a given state with respect to two or more incompatible bases. The lower bounds on the sum of the relative entropies of coherence for single and bipartite quantum systems are facilitated through the use of entropic uncertainty relations with and without memory effects. Thus, entropic uncertainty relations may be viewed as restrictions on coherence of a quantum system in two incompatible bases. For bipartite quantum systems, we find an upper bound on the sum of the relative entropies of coherence via Lewenstein-Sanpera decomposition \cite{Lewenstein98} of bipartite quantum states. We then show that this upper bound can be tightened.  We also provide a non trivial upper bound on the absolute value of the differences between the relative entropy of coherence obtained in two different bases.

Let us formally introduce the measures of coherence in the framework of resource theory, that is based on the set of incoherent operations and incoherent states.  First we fix a reference basis and choose it to be the computational basis $\{\ket{i}\}$ (without loss of any generality). A maximally coherent state in this basis is given by $\ket{\psi_d} = \frac{1}{\sqrt{d}}\sum_{i=0}^{d-1} \ket{i}$ as any other state can be created from $\ket{\psi_d}$ using only the set of incoherent operations. Also, the coherence of this state naturally provide a reference to gauge the coherence of other quantum states. The set of all the states of the form $\rho_I= \sum_{i}d_i \ketbra{i}{i}$, where $d_i\geq0$ and $\sum_i d_i=1$, forms the set $\mathcal{I}$ of all incoherent states. Any quantum state that does not belong to the set $\mathcal{I}$ will be called a coherent state and will act as a resource. 
In quantum theory, a physical operation is represented by a completely positive trace preserving (CPTP) map. Any CPTP map can be expressed in terms of a set of Kraus operators $\{M_k\}$ such that the state $\rho$ of the system is transformed to $\rho\rightarrow\sum_kp_k\rho_k$, where $\rho_k = M_k \rho M^\dagger_k/p_k$ and $p_k = \mathrm{Tr}[M_k \rho M^\dagger_k]$. An operator is said to be an incoherent operator if it maps diagonal states to some other diagonal states in the reference basis. Any quantum channel whose Kraus elements are the incoherent operators for a reference basis is said to be an incoherent channel (ICPTP map) and such an ICPTP map $\Lambda_I$ satisfies $\Lambda_I[\mathcal{I}]\subseteq \mathcal{I}$.
In the following, we list the conditions for a real valued function of quantum states that make the function a bona fide measure of quantum coherence. Following Ref. \cite{Baumgratz2014}, a~function on the quantum state $C^{(i)}(\rho)$ (the superscript $i$ is used to denote the reference basis) is said to be a valid measure of quantum coherence of the state $\rho$ if it satisfies following conditions:
(1) $C^{(i)}(\rho) = 0$ iff $\rho \in \mathcal{I}$. (2) $C^{(i)}(\rho)$ is non increasing under the incoherent operations, i.e., for any incoherent channel denoted by $\Lambda_I$, we have $C^{(i)}(\Lambda_I[\rho]) \leq C^{(i)}(\rho)$. (3) $C^{(i)}(\rho)$ decreases on an average under the selective incoherent
operations, i.e., $\sum _k p_k  C^{(i)}(\rho_k) \leq C^{(i)}(\rho)$,
where $\rho_k = M_k \rho M^\dagger_k/p_k$, $p_k = \mathrm{Tr}[M_k \rho M^\dagger_k]$ and $M_k$ are the Kraus elements of an incoherent
channel. (4) $C^{(i)}(\rho)$ is a convex function of quantum states, i.e., $C^{(i)}(\sum_k p_k \rho_k) \leq \sum_k p_k C^{(i)}(\rho_k)$. It~may be noted that the conditions (3) and (4) together imply condition (2). The bona fide measures of coherence include the relative entropy of coherence, the $l_1$ norm of coherence \cite{Baumgratz2014}, the coherence of formation and the distillable coherence \cite{Winter2015}. In this work, we will work only with the relative entropy of coherence, which for a density matrix $\rho$ and a reference basis $\{\ket{i}\}$, is defined as
\begin{align}
C^{(i)}(\rho) = S(\rho_d^{(i)}) - S(\rho)
\end{align}
where $S(\rho) = -\mathrm{Tr}[\rho \log \rho]$ is the von Neumann entropy. Here and in the rest of the paper, all the logarithms are taken with respect to base $2$. The superscript $i$ in $C^{(i)}(\rho)$ indicates that coherence is calculated in reference basis $\{\ket{i}\}$. If~$\rho$ is a pure density operator and $\rho=\ket{\psi}\bra{\psi}$, then we have $C^{(i)}(\rho) = S(\rho_d^{(i)})$, where $\rho_d^{(i)}$ is the diagonal part of $\rho=\ket{\psi}\bra{\psi}$ in the basis $\{\ket{i}\}$.

\section{Uncertainty Like Relation for Quantum Coherence Expressed in Two Different Non-Commuting~Bases}
Let us first consider the relation between the coherence of a single $d$-level quantum system in a state $\rho$ expressed in two different non commuting bases. Let the sets $P = \{\ket{i}\bra{i}\}$ and $Q=\{\ket{a}\bra{a}\}$ denote two non-commuting projective measurements on the given quantum system. The state of the system after the projective measurements by $P$ and $Q$, is given, respectively, by
\begin{align}
 &\rho_P = \sum_{i} p_i\ketbra{i}{i}~~\mathrm{and}\\
 &\rho_Q = \sum_{a} q_a\ketbra{a}{a}
\end{align}
where $p_i = \bra{i}\rho\ket{i}$ and $q_a = \bra{a}\rho\ket{a}$. Now the entropic uncertainty relation \cite{Maassen1988, Berta2010} for these two measurements~reads
\begin{align}
\label{u1}
 H(P) + H(Q) \geq -2\log C + S (\rho)
\end{align}
where $H(P) = -\sum_i p_i\log p_i$, $H(Q) = -\sum_a q_a\log q_a$, $S(\rho) = -\mathrm{Tr}[\rho \log \rho]$ and $C = \max_{i,a}|\bra{i} a\rangle|$. Also, $H(P) = S(\rho_d^{(i)})$ and $H(Q) = S(\rho_d^{(a)})$, where $\rho_d^{(i)}$ and $\rho_d^{(a)}$ are the diagonal parts of the density matrix $\rho$ in bases $\{\ket{i}\}$ and $\{\ket{a}\}$, respectively. Now, Equation (\ref{u1}) can be rewritten as
\begin{align}
 &C^{(i)}(\rho) + C^{(a)}(\rho) \geq -2\log C - S (\rho)
\end{align}
where $C^{(i)}(\rho) = S(\rho_d^{(i)}) - S(\rho)$ and $C^{(a)}(\rho) = S(\rho_d^{(a)}) - S(\rho)$ are the coherences of the density matrix $\rho$ in the bases $\{\ket{i}\}$ and $\{\ket{a}\}$, respectively. The above relation can be termed as the uncertainty relation for the coherences of a density matrix measured in two different (non-commuting) bases. Since the coherence of a quantum system is a
basis dependent notion, above relation gives us insight about the coherences of a same density matrix measured in two different bases.

Now let us consider a bipartite quantum system in a state $\rho_{AB}$. Also, let $\rho_B = \mathrm{Tr}_A[\rho_{AB}] = \sum_{\mu}\lambda_{\mu}\ketbra{\mu}{\mu}$, where $\lambda_\mu$ are the eigenvalues and $\ket{\mu}$ are the eigenvectors of $\rho_B$. Consider two projective measurements on the state $\rho_{AB}$, given by $P_{AB} = \{\ketbra{i}{i}\otimes \ketbra{\mu}{\mu} \}$ and $Q_{AB} = \{\ketbra{a}{a}\otimes \ketbra{\mu}{\mu} \}$. Note that these projective measurements do not disturb the density operator $\rho_B$. The state of the bipartite system after the projective measurements by $P_{AB}$ and $Q_{AB}$, is given, respectively, by
\begin{align}
 &\rho_{PR} = \sum_{i,\mu} p_{i\mu}\ketbra{i}{i}\otimes \ketbra{\mu}{\mu} \mathrm{~~and}\\
 &\rho_{QT} = \sum_{a,\mu} q_{a\mu}\ketbra{a}{a}\otimes \ketbra{\mu}{\mu}
\end{align}
where $p_{i\mu} = \bra{i,\mu}\rho_{AB}\ket{i,\mu}$ and $q_{a\mu} = \bra{a,\mu}\rho_{AB}\ket{a,\mu}$. Noting that the projective measurements always increase entropy, we have $S(\rho_{PR})\geq S(\rho_{PB})$ and similarly, $S(\rho_{QT})\geq S(\rho_{QB})$. Here, $\rho_{PB}=\sum_i \ketbra{i}{i} \otimes \mathbb{I} \rho_{AB} \ketbra{i}{i} \otimes \mathbb{I}$ and similarly $\rho_{QB}=\sum_a \ketbra{a}{a} \otimes \mathbb{I} \rho_{AB} \ketbra{a}{a} \otimes \mathbb{I}$.
Also note that $S(\rho_R) = S(\rho_T) = S(\rho_B)$. Therefore, $H(P|R) \geq H(P|B)$ and $H(Q|T) \geq H(Q|B)$, where $H(X|Y) = S(\rho_{XY}) - S(\rho_Y)$. This gives
\begin{align}
\label{u2}
 H(P|R) + H(Q|T) &\geq H(P|B) + H(Q|B)\nonumber\\ 
 &\geq -2\log C + S (A|B)
\end{align}
where the last inequality follows from entropic uncertainty relation in the presence of memory, first given in the Ref. \cite{Berta2010}, and $C = \max_{i,a}|\bra{i} a\rangle|$. Now $H(P|R) = S(\rho_{PR}) - S(\rho_B) =  C^{(i\mu)}(\rho_{AB}) + S(\rho_{AB}) - S(\rho_B) = C^{(i\mu)}(\rho_{AB}) + S(A|B)$. Similarly, $H(Q|T) =  C^{(a\mu)}(\rho_{AB}) + S(A|B)$. Therefore, Equation~(\ref{u2})~becomes
\begin{align}
\label{eq:coh}
 C^{(i\mu)}(\rho_{AB}) + C^{(a\mu)}(\rho_{AB}) \geq -2\log C - S (A|B)
\end{align}
For pure bipartite entangled states $S(A|B)$ is negative, and hence the above uncertainty like relation for quantum coherence is tightened. This is in contrast to the memory assisted uncertainty relation, where entanglement reduces the uncertainty. Therefore, in the presence of entanglement, quantum coherence and entropic uncertainty relation seemingly behave in opposite ways. The entropic uncertainty relation in the presence of quantum memory has been improved in Ref. \cite{Arun2012} and using it, we have
\begin{align}
H(P|R) + H(Q|T) &\geq H(P|B) + H(Q|B)\nonumber\\ 
 &\geq -2\log C + S (A|B) - \max\{0,\mathcal{D}-\mathcal{J}\}\nonumber
\end{align}
where $\mathcal{D}$ is quantum discord across $AB$ partition \cite{Henderson2001, Ollivier2001, Modi2012} and $\mathcal{J}$ is the classical correlation \cite{Henderson2001, Ollivier2001, Modi2012}. Therefore the Equation (\ref{eq:coh}) is improved to
\begin{align}
\label{eq:dis-bound}
 C^{(i\mu)}(\rho_{AB}) + C^{(a\mu)}(\rho_{AB}) \geq -2\log C - S (A|B)- \max\{0,\mathcal{D}-\mathcal{J}\}
\end{align}
If $\mathcal{D}<\mathcal{J}$, the bound in Equation (\ref{eq:dis-bound}) becomes even more tighter.

For a tripartite state $\rho_{ABD}$, one may ask how does the entanglement across $AB$ and $AD$ affects the uncertainty like relation for quantum coherence. Similar to Equation (\ref{eq:coh}), one can have a uncertainty like relation for quantum coherence for the density operator $\rho_{AD}$, which is given by
 \begin{align}
 C^{(i\nu)}(\rho_{AD}) + C^{(a\nu)}(\rho_{AD}) \geq -2\log C - S (A|D)
\end{align}
It turns out that entanglement cannot tighten uncertainty like relation for quantum coherence across $AB$ and $AD$ at the same time. This is because, we have $S(A|B) + S(A|D) \ge 0$ for any tripartite quantum state, which follows from the strong subadditivity of the von Neumann entropy \cite{Nielsen10, Hu2013}. This shows that whenever we have $S(A|B) < 0$ , we have $S(A|D) > 0$. This means that if one tightens the uncertainty like relation for quantum coherence across $AB$ partition, the uncertainty like relation for quantum coherence across $AD$ partition cannot be improved despite the presence of entanglement in the state $\rho_{ABD}$.
 
\section{Uncertainty Like Relation for Quantum Coherence Expressed in Many Different Non-Commuting~Bases}
Let us first consider a single $d$-level quantum system in a state $\rho$. Let the sets $P_1 = \{\ket{i_1}\bra{i_1}\}$ and $P_2=\{\ket{i_2}\bra{i_2}\},\ldots,P_n =\{\ket{i_n}\bra{i_n}\} $ denote $n$ non-commuting projective measurements on the given quantum system. The state of the system after the projective measurements by $P_1,\ldots,P_n$ is given, respectively, by~$ \rho_{P_1} = \sum_{i_1} p_{i_1}\ketbra{i_1}{i_1},\ldots,\rho_{P_n} = \sum_{i_n} p_{i_n}\ketbra{i_n}{i_n}$, where $p_{i_1} = \bra{i_1}\rho\ket{i_1},\ldots,p_{i_n} = \bra{i_n}\rho\ket{i_n}$. Now the entropic uncertainty relation \cite{HFan14} for multiple measurements read
\begin{align}
\label{u11}
 \sum_{i=1}^{n} H(P_i) \geq -\log b + (n-1)S (\rho)
\end{align}
where $H(P_k) = -\sum_{i_k} p_{i_k}\log p_{i_k}$, $(k=1,\ldots,n)$, $S(\rho) = -\mathrm{Tr}[\rho \log \rho]$ and
\begin{align}
\label{b}
 b = \max_{i_n}\left\{\sum_{i_2,.,i_{n-1}}\max_{i_1} [c(i_1,i_2)]\prod_{k=2}^{n-1}c(i_k,i_{k+1}) \right\}
\end{align}
where $c(i_k,j_l) = |\langle i_k|j_l \rangle|^2$. Also, $H(P_k) = S(\rho_d^{(i_k)})$ with $k=1,\ldots,n$, where $\rho_d^{(i_k)}$ are the diagonal parts of the density matrix $\rho$ in bases $\{\ket{i_k}\}$. Now, Equation (\ref{u11}) can be rewritten as
\begin{align}
 &\sum_{k=1}^{n}S(\rho_d^{(i_k)}) \geq -\log b + (n-1)S (\rho), ~\mathrm{or}\nonumber\\
 &\sum_{k=1}^{n}C^{(i_k)}(\rho) \geq -\log b - S (\rho)
\end{align}
where $C^{(i_k)}(\rho) = S(\rho_d^{(i_k)}) - S(\rho)$ is the coherences of the density matrix $\rho$
in the $\{\ket{i_k}\}$ basis.

Now let us consider a bipartite quantum system in a state $\rho_{AB}$, where $\rho_B = \mathrm{Tr}_A[\rho_{AB}] = \sum_{\mu}\lambda_{\mu}\ketbra{\mu}{\mu}$, and $n$ projective measurements on the state $\rho_{AB}$, given by $P^i_{AB} = \{\ketbra{i_k}{i_k}\otimes \ketbra{\mu}{\mu} \}$ with $k=1,\ldots,n$. The state of the bipartite system after the projective measurements by $P^k_{AB}$ is given~by
\begin{align}
 \rho_{P_kR_k} = \sum_{i_k,\mu} p_{i_k\mu}\ketbra{i_k}{i_k}\otimes \ketbra{\mu}{\mu}
\end{align}
where $p_{i_k\mu} = \bra{i_k,\mu}\rho_{AB}\ket{i_k,\mu}$. Noting that the projective measurements
always increase entropy, we~have $S(\rho_{P_kR_k})\geq S(\rho_{P_kB})$, with $\rho_{P_kB}=\sum_{i_k} \ketbra{i_k}{i_k} \otimes \mathbb{I} \rho_{AB}  \ketbra{i_k}{i_k} \otimes \mathbb{I}$. Also note that
$S(\rho_{R_k}) = S(\rho_B)$. Therefore, $H(P_k|R_k) \geq H(P_k|B)$. Now,
\begin{align}
\label{u22}
 \sum_{k=1}^{n} H(P_k|R_k) &\geq \sum_{k=1}^{n} H(P_k|B)\nonumber\\ 
 &\geq -\log b + (n-1)S (A|B)
\end{align}
where the last inequality follows from the many measurement generalization of entropic uncertainty relation in presence of memory \cite{Berta2010} and is obtained in Ref. \cite{HFan14}, and $b$ is given by Equation~(\ref{b}).
Now~$H(P_k|R_k) = S(\rho_{P_kR_k}) - S(\rho_B) =  C^{(i_k\mu)}(\rho_{AB}) + S(\rho_{AB}) - S(\rho_B) = C^{(i_k\mu)}(\rho_{AB}) + S(A|B)$.
Therefore,~Equation (\ref{u22}) becomes
\begin{align}
 \sum_{k=1}^{n}C^{(i_k\mu)}(\rho_{AB}) \geq -\log b - S (A|B)
\end{align}
For bipartite pure entangled states, the above inequality is tightened. 

\section{Complementarity Like Relation for Quantum Coherence for a Bipartite State}

For a bipartite quantum system, there exists an optimal
decomposition of every density matrix $\rho_{AB}$ of the bipartite system, called as the
Lewenstein-Sanpera (LS) decomposition, in terms of a separable state and an 
entangled state \cite{Lewenstein98, Karnas01, Jafarizadeh04, Englert09, Thiang10}, i.e.,
\begin{align}
\label{LS}
 \rho_{AB} = \lambda \rho_s + (1-\lambda)\rho_e;~\lambda \in [0,1]
\end{align}
where $\rho_s$ is a separable state on the separable-entangled boundary and $\rho_e$ is an entangled state. The~parameter $\lambda$ is taken to be optimal in the sense that any other decomposition of the form $\lambda' \rho'_s + (1-\lambda')\rho'_e$ with $\lambda' \in [0,1]$ and $\rho_s\neq \rho'_s$ necessarily implies $\lambda'< \lambda$. For optimal $\lambda$, $\rho_s$ is called the optimal separable approximation (OSA) of the state $\rho_{AB}$. For a bipartite qubit system the optimal LS decomposition becomes unique with the entangled part of the decomposition being a pure entangled projector \cite{Lewenstein98}. Using the concavity of the von Neumann entropy, Equation (\ref{LS}) implies
$S(\rho_{AB}) \geq \lambda S(\rho_s)$. Let us denote the quantum coherence of the bipartite state $\rho_{AB}$ in two different bases $\{\ket{ij}\}$ and $\{\ket{ab}\}$ as $C^{ij}(\rho_{AB})$ and $C^{ab}(\rho_{AB})$, respectively, then
\begin{align}
\label{eq:old-rhs1}
 C^{ij}(\rho_{AB}) + C^{ab}(\rho_{AB}) &= S(\rho^{ij}_d) +  S(\rho^{ab}_d) -  2 S(\rho_{AB})\nonumber\\
 &\leq 2 \log d_A d_B - 2K(\rho_{AB})
\end{align}
where $K(\rho_{AB}) = \lambda S (\rho_s) + (1-\lambda) S (\rho_e)$ and $d_{A(B)}$ being the dimension of the system $A(B)$. This~tightens the trivial upper bound on the sum of coherences, namely $2 \log d_A d_B$. For bipartite qubit systems, since the entangled part in the optimal separable approximation becomes pure, $K(\rho_{AB})$~becomes $\lambda S(\rho_s)$ and we have
\begin{align}
 C^{ij}(\rho_{AB}) + C^{ab}(\rho_{AB}) \leq 4- 2\lambda S (\rho_s)
\end{align}

\begin{figure}[htbp!]
\centering
\subfigure[~$d_1=0.52$, $d_2=0.1$]
{
\includegraphics[width=70 mm]{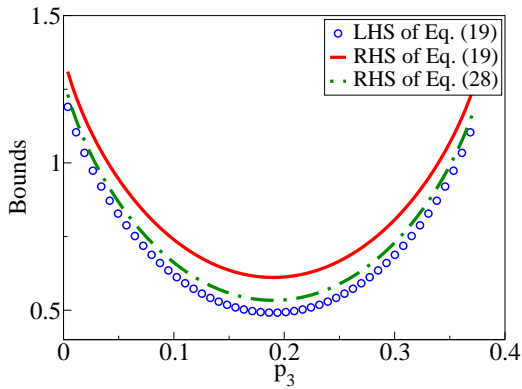}
}
\centering
\subfigure[~$d_1=0.6$, $d_2=0.05$]
{
\includegraphics[width=70 mm]{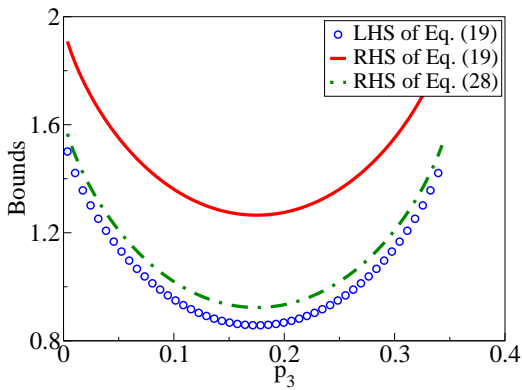}
}
\caption{The plot shows the left hand and right hand sides of the uncertainty like relations for quantum coherence, given by Equations (\ref{eq:old-rhs1}) and (\ref{eq:new-rhs}), for Bell diagonal states. It clearly shows that the bound in Equation~(\ref{eq:new-rhs}) is tighter than that in Equation (\ref{eq:old-rhs1}). (\textbf {a}) $d_1=0.52$, $d_2=0.1$; (\textbf {b})~$d_1=0.6$, $d_2=0.05$.}
\label{fig:figure1}
\end{figure}

In the following, we provide some examples of optimal LS decomposition and calculate the bound
on the sum of the coherences in two different bases. Consider a $2\otimes 2$ Bell diagonal state
\begin{align}
 \rho_{AB} = \sum_{i=1}^{4}d_i\ket{B_i}\bra{B_i}
\end{align}
where $\ket{B_1} = (\ket{00}+\ket{11})/\sqrt{2}$, $\ket{B_2} = (\ket{00}-\ket{11})/\sqrt{2}$, $\ket{B_3} = (\ket{01}+\ket{10})/\sqrt{2}$, $\ket{B_4} = (\ket{01}-\ket{10})/\sqrt{2}$ and $\{d_i\}$ is a probability vector. The optimal LS decomposition, $\rho_{AB} = \lambda \rho_s + (1-\lambda) \rho_e$ of the above Bell diagonal state is obtained for $\rho_e = \ket{B_1}\bra{B_1}$ and $\rho_s = \sum_{i=1}^{4}d'_i\ket{B_i}\bra{B_i}$ where $d'_1 = 1/2$, $d'_j = d_j/\lambda$ $(j=2,3,4)$ and $\lambda = 1 - \mu$ \cite{Jafarizadeh04}. Here, $\mu$ is the concurrence \cite{Wooters97, Wootters98} of the initial state
$\rho_{AB}$. The quantity $K(\rho_{AB})$, here, is given by $\frac{(1-\mu)}{2}\log 2 + (1-d_1)\log (1-\mu)-\sum_{i=2}^{4}d_i \log d_i$. For a specific example, take $d_1=0.6$, $d_2=0.2$, $d_3=0.1=d_4$. In this case $\mu = 0.2$ and $K(\rho_{AB})=1.4$, where $\log$ is calculated using base $2$. Therefore, sum of the coherences for the Bell diagonal state, given by above parameters, satisfy
\begin{align}
 C^{ij}(\rho_{AB}) + C^{ab}(\rho_{AB}) \leq 1.2
\end{align}
This bound gives a value which is well below the trivial bound which is four. Another example that we consider is the case of bipartite qutrit state which contains bound entanglement \cite{Horodecki1999} in it. The~state is given by
\begin{align}
\label{state}
 \rho = \frac{2}{7}\ketbra{\psi_+}{\psi_+} + \frac{\gamma}{7}P_+ + \frac{5-\gamma}{7}P_-;~~~2\leq \gamma \leq 5
\end{align}
where $\ket{\psi_+} = (\ket{00}+\ket{11}+\ket{22})/\sqrt{3}$, $P_+ = (\ket{01}\bra{01}+\ket{12}\bra{12}+\ket{20}\bra{20})/3$ and $P_- = (\ket{10}\bra{10}+\ket{21}\bra{21}+\ket{02}\bra{02})/3$. The optimal LS decomposition, $\rho = \lambda \rho_s + (1-\lambda \rho_e)$ of the state $\rho$ is obtained for $\rho_e = \frac{2}{7}\ketbra{\psi_+}{\psi_+} + \frac{5}{7}P_+ $, $\rho_s = \frac{2}{7}\ketbra{\psi_+}{\psi_+} + \frac{3}{7}P_+ +  \frac{2}{7}P_- $ and $\lambda = (5-\gamma)/2$ \cite{Jafarizadeh04}. The~quantity $K(\rho_{AB})$, here, is given by $0.8631+0.6935\lambda$. Therefore, sum of the coherences for the state $\rho$, satisfy
\begin{align}
\label{cp3}
 C^{ij}(\rho_{AB}) + C^{ab}(\rho_{AB}) \leq 2\log 9- 2(0.8631+0.6935\lambda)
\end{align}
The bound, Equation (\ref{eq:old-rhs1}), can be tightened as follows. Let us consider a bipartite state $\rho_{AB}$. Also, assume that the marginal density matrix $\rho_B$ for subsystem $B$ is diagonal in the basis $\ket{\mu}$, i.e., $\rho_B = \sum_\mu d_\mu \ketbra{\mu}{\mu}$. Now let us consider a projective measurement on the bipartite system in the basis $\{\ket{i}\otimes\ket{\mu}\}$. The state of the total system after the measurement, is given by
\begin{align}
 \rho_{PR} = \sum_{i,\mu}\bra{i,\mu}\rho_{AB}\ket{i,\mu}\ketbra{i,\mu}{i,\mu}
\end{align}
The marginal density matrix $\rho_R$ of the subsystem $B$ after the measurement is same as $\rho_B$.
Now, $
 H(P|R) = H(PR)-H(R) = S(\rho_{PR}) - S(\rho_B)= C^{(i\mu)}(\rho_{AB}) + S(A|B)$.
Therefore,
\begin{align}
 C^{(i\mu)}(\rho_{AB}) &= H(P|R)-S(A|B) = H(P) - I(P:R) - S(A|B)\nonumber\\
 &\leq H(P) - S(A|B)\nonumber\\
 &\leq \log d_A - S(A|B)
\end{align}
where $d_A$ is the dimension of the subsystem $A$. Now, if we consider the projective measurement of the total system in the basis $\{\ket{a,\mu}\}$, we get
\begin{align}
 C^{(a\mu)}(\rho_{AB}) \leq \log d_A - S(A|B)
\end{align}
Therefore,
\begin{align}
\label{eq:new-rhs}
 C^{(i\mu)}(\rho_{AB}) + C^{(a\mu)}(\rho_{AB}) \leq 2\log d_A - 2S(A|B)
\end{align}
Again we consider the example of the Bell diagonal state 
$ \rho_{AB} = \sum_{i=1}^{4}d_i\ketbra{B_i}{ B_i}$,
where $\ket{B_1} = (\ket{00}+\ket{11})/\sqrt{2}$, $\ket{B_2} = (\ket{01}+\ket{10})/\sqrt{2}$, $\ket{B_3} = (\ket{01}-\ket{10})/\sqrt{2}$, and $\ket{B_4} = (\ket{00}+\ket{11})/\sqrt{2}$. \linebreak For~$d_1>1/2$,~the optimal LS decomposition is given by: $\lambda = 2(1-d_1)$, $\rho_e=\ketbra{B_1}{B_1}$, and 
\begin{align}
 \rho_s = \sum_{i=1}^{4}d'_i\ketbra{B_i}{ B_i}
\end{align}
where $d'_1 = 1/2$, $d'_j = d_j/\lambda$ $(j=2,3,4)$. Note that the previous bound in Equation (\ref{eq:old-rhs1}) is given by
\begin{align}
\label{old}
 &2\log d_Ad_B - \lambda S(\rho_s)\nonumber\\
 &=4 - (1-d_1)(2+\log [1-d_1]) + \sum_{i=2}^{4}d_i \log d_i
\end{align}
However, the bound in Equation (\ref{eq:new-rhs}) is given by
\begin{align}
\label{eq:new-rhs1}
2\log d_A - 2S(A|B) &= 2 - 2S(\rho_{AB}) + 2S(B)\nonumber\\
 & = 4 + 2\sum_{i=1}^{4}d_i \log d_i
\end{align}
For Bell diagonal states with $d_1>1/2$, the new bound (Equation (\ref{eq:new-rhs})) is always smaller than the bound (Equation (\ref{eq:old-rhs1})) (see also Figure \ref{fig:figure1}).

\section{State Dependent Upper Bound for Coherence}
One may ask that if diagonal parts of a density operator in two different reference bases are close, does this mean the closeness of coherences of the density operator in the given two reference bases?
Let us consider a quantum system in a state $\rho$, with $\rho^{(i)}_d$ being the diagonal part of
the density matrix in $\{\ket{i}\}$ basis.  For, $||\rho^{(i)}_d - \rho||_1 = 2\epsilon$,
with $||.||_1$ being the trace norm, defined as $||M||_1 = \mathrm{Tr}\sqrt{M^\dagger M}$, the 
Fannes-Audenaert inequality \cite{Fannes73,Audenaert07} implies
\begin{align}
 C^{(i)}(\rho) = |S(\rho^{(i)}_d) - S(\rho)| \leq \epsilon \log (d-1) + H (\epsilon)
\end{align}
where $d$ is the dimension of the Hilbert space of the quantum system and $H(\epsilon)$ is the binary entropy. Let the diagonal part
of the density matrix $\rho$ in the basis $\{\ket{a}\}$ be denoted by
$\rho^{(a)}_d$. For $||\rho^{(i)}_d - \rho^{(a)}_d||_1 = 2\eta$, again using 
the Fannes-Audenaert inequality \cite{Fannes73,Audenaert07,Nielsen10}, we have
\begin{align}
\label{af1}
 |S(\rho^{(i)}_d) - S(\rho^{(a)}_d)| \leq \eta \log (d-1) + H (\eta)
\end{align}
Note that $S(\rho^{(i)}_d) - S(\rho^{(a)}_d) = [S(\rho^{(i)}_d)- S(\rho)] - [S(\rho^{(a)}_d)- S(\rho)] =C^{(i)}(\rho) - C^{(a)}(\rho) $.
Now, Equation~(\ref{af1})~implies
\begin{align}
 |C^{(i)}(\rho) - C^{(a)}(\rho)| \leq \eta \log (d-1) + H (\eta)
\end{align}
When $\eta=||\rho^{(i)}_d - \rho^{(a)}_d||_1/2$ is small, i.e., the diagonal parts of a quantum state in two different  bases are close to each other with respect to the trace distance, the relative entropies of coherence of the state in these two bases are also close to each other.

\section{Summary}
In quantum theory, uncertainty relation is a fundamental consequence of superposition principle and incompatible nature of observables. As quantum coherence is a basis dependent notion, it is pertinent to ask if coherence respects some kind of uncertainty relation for two or more incompatible bases.
In this paper, we have explored the interplay of the relative entropy of coherence of a quantum system in a given state in two or more incompatible bases. We have proved trade-off relations for the relative entropy of coherence in two or more non-commuting bases for single and bipartite quantum systems. This shows that the relative entropies of coherence of a quantum system in two or more incompatible bases are not independent of one another. If, in one basis, the density matrix shows a larger value of the relative entropy of coherence, in another basis it may not show the same value. In~the case of bipartite states, the presence of entanglement tightens the trade-off relation for the relative entropy of coherence. However, because of strong subadditivity of conditional entropy, one cannot have tightened trade-off relations for quantum coherence measure across two different parties. Also, we have proved complementarity like relations for the relative entropy of coherence in two different bases.
Moreover, we have provided an upper bound on the absolute value of the differences of the relative entropy of coherence obtained in two different bases.

\begin{acknowledgments}
M.N.B. acknowledges financial support from the John Templeton Foundation, the Spanish MINECO (FIS2013-46768, FIS2008-01236, and FIS2013-40627-P) with the support of FEDER funds, ``Severo Ochoa'' Programme (SEV-2015-0522), the Generalitat de Catalunya (2014-SGR-874 and 2014-SGR-966), and Fundaci\'{o} Privada Cellex. U.S.  acknowledges the research fellowship of Department of Atomic Energy, Government of India.
\end{acknowledgments}

\bibliographystyle{apsrev4-1}
\bibliography{literature}

\end{document}